\documentclass[11pt,a4paper]{article}
\usepackage[top=22mm,bottom=22mm,left=22mm,right=22mm]{geometry}
\usepackage[T1]{fontenc}
\usepackage[utf8]{inputenc}
\usepackage{amsmath,amssymb,bm,graphicx,microtype,booktabs}
\usepackage[numbers,sort&compress]{natbib}
\usepackage[hidelinks]{hyperref}
\usepackage[font=small,labelfont=bf]{caption}
\hypersetup{pdftitle={Matter that learns what matters},pdfauthor={N. G. Berloff}}
\graphicspath{{figures/}}
\title{\bfseries Matter that learns what matters}
\author{N.~G.~Berloff\\[3pt]
\small Department of Applied Mathematics and Theoretical Physics,\\
\small University of Cambridge, Cambridge CB3 0WA, United Kingdom}
\date{}
\begin{document}
\raggedbottom
\maketitle
\begin{abstract}
\noindent
Can a material learn what deserves its attention? We show this in a model optical network, where action outcomes train a persistent record in each link. The record biases whether a source is consulted and sets how strongly it changes a stored answer, so useful sources gain attention and influence without being identified in advance. The network's experience thus shapes the information available for its future learning. Resonant readouts make this learning fragile: reinforcement can carry a record past resonance, where it persists but its influence weakens. We derive bounds within which relevant signals guide the answer, irrelevant signals are rejected and stored answers remain open to revision. These bounds guide a learning rule that limits reinforcement as records form; in simulations, it acquires useful records and keeps them functional. When relevance changes, an obsolete record can block its own replacement. A local state that strengthens negative writing after unsuccessful actions releases obsolete records and completes the learning sequence more often. A reduced stochastic model, calibrated retrospectively from local response statistics, accounts for changes in sequence completion across writing rules without fitting the aggregate success rates. How a memory is read should decide how it is written.
\end{abstract}

An animal searching for food hears many sounds, but only a few tell it where to look. Experience teaches it which signals to examine and how strongly an observation should change what it remembers. This is learned attention: the selection and weighting of information~\cite{mackintosh1975,pearcehall1980,gottlieb2013}. Which signals are useful can also change. A source that once guided successful action may later mislead. A learning system therefore needs both a way to favour useful information and a way to revise that preference. Can a material acquire this control through experience?

Physical systems retain records of past forcing~\cite{keim2019} and can learn responses~\cite{kaspar2021,bhattacharyya2022,stern2021,stern2023,dillavou2022,scellier2017,lopezpastor2023,wright2022,momeni2023,ibrahim2026,du2026}. Such learning can use local rules~\cite{dillavou2024,feldmann2019} and reward~\cite{loeffler2023,dillavou2025}. Optical states can also route light~\cite{lugnan2025}, while reward-trained optical reservoirs have used off-chip readouts~\cite{kanno2020,kanno2022}. Here, action outcomes train a persistent material state that governs both which source is consulted and how strongly it changes memory. We call this \emph{endogenous information control}. The network's experience shapes the evidence available for its future learning. Our use of attention refers to this learned control of information, rather than transformer attention.

This creates a feedback loop. The learned record helps determine which observations can strengthen or revise it. A useful preference can guide further learning, but an obsolete preference can restrict access to the evidence needed for correction.

\begin{figure}[p]
\centering
\includegraphics[width=0.90\linewidth]{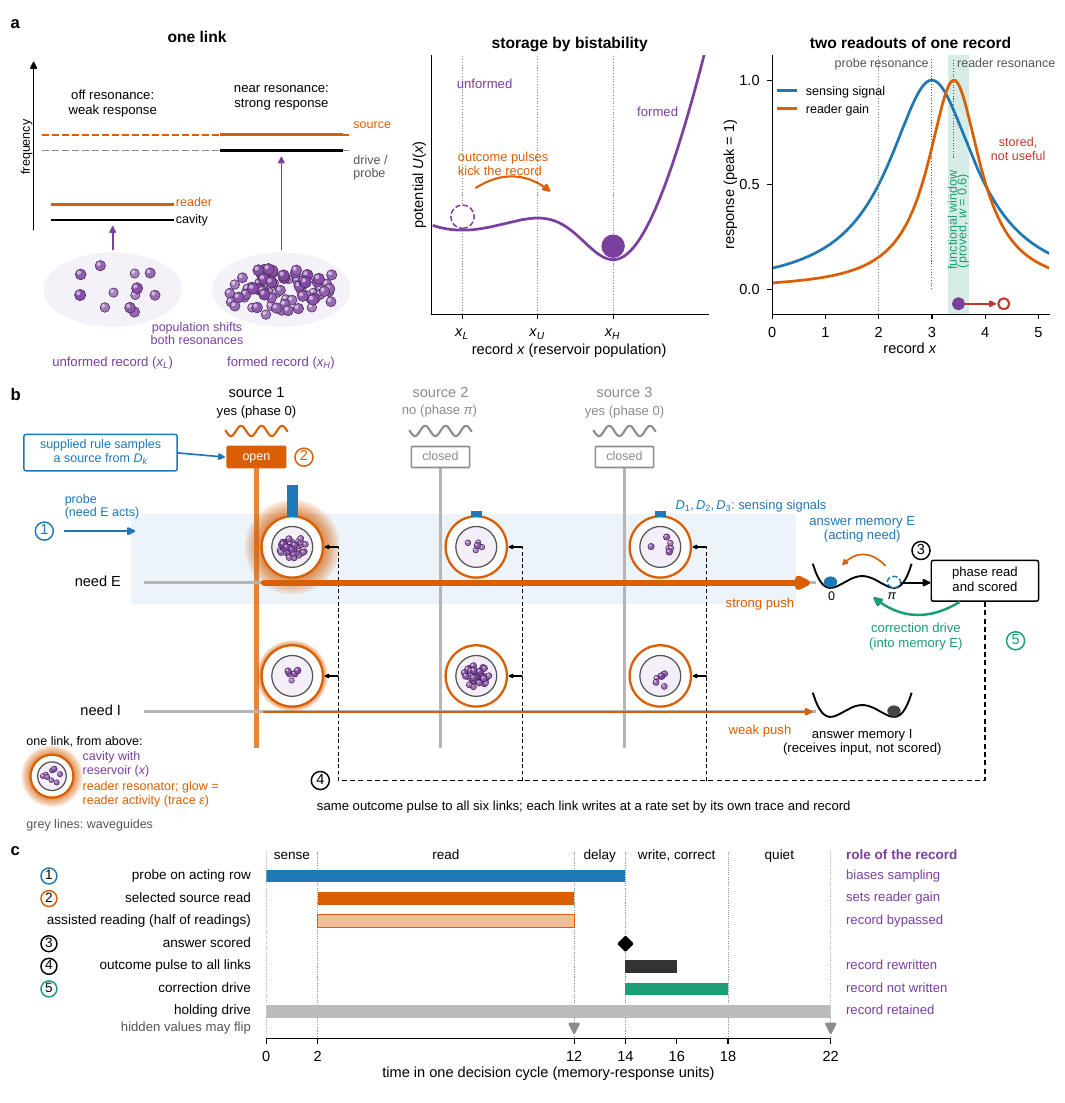}
\caption{\textbf{One population per link controls both source choice and memory revision.}
\textbf{a}, One link. Left: absorbed light fills a reservoir of excitations (purple balls). Its population $x$, the relevance record, shifts the cavity and reader resonances relative to the fixed frequencies of the drive, probe and source field.
Middle: optical bistability, Eq.~(\ref{eq:red}), shown as motion in the effective potential $U(x)$. Its minima are the stable populations of the unformed ($x_L$, dashed circle) and formed ($x_H$, filled circle) records. Outcome pulses can push the record across the barrier at $x_U$.
Right: the two readouts, normalised to their peaks. The cavity's probe response $1/D(x)$ is proportional to the sensing signal and biases source choice. The reader gain $g(x)/g_0$, shown for width $w=0.6$, sets how strongly the source moves the answer. Both peak near resonance. Green marks the range proved at $w=0.6$, with valid records in $[3.30,3.70]$, where both readouts work (the functional window, Fig.~\ref{fig:window}). Beyond this range (arrow), a record can stay formed but cease to be useful (open circle).
\textbf{b}, Network crossbar. Source lines (columns; only the selected one is open) cross memory lines (rows, one per need). At each crossing, a cavity holds a reservoir of population $x_{jk}$, and a reader resonator couples the lines efficiently only near resonance. The reader's glow represents the activity recorded by the trace $\epsilon_{jk}$ under the admission rule.
(1)~The probe of the acting need~E gives the sensing signals $D_k$ (blue bars): cavity intensities proportional to $1/D(x)$ in panel a.
(2)~A supplied rule, Eq.~(\ref{eq:selection}), selects a source. It pushes memory~E strongly through the formed link and memory~I weakly through the unformed one. The source reaches both memories, but the records determine which memory it can change.
(3)~The phase of memory~E is scored.
(4)~One outcome pulse, a proposed device operation, reaches all six links; each writes at a rate $\Pi_{jk}\propto\epsilon_{jk}H(x_{jk})$, Eq.~(\ref{eq:writer}).
(5)~The correction drive resets memory~E; it writes no record.
\textbf{c}, Timing of one decision cycle and the role of the record in each step; assisted reading bypasses the record of the selected reader.}
\label{fig:cycle}
\end{figure}

We make the question concrete in a model optical network (Fig.~\ref{fig:cycle}; Table~\ref{tab:glossary}). It keeps an answer to each of two yes-or-no questions about hidden values that can change. Each question, or \emph{need} (E or I), has an optical \emph{answer memory} whose two stable phases represent yes and no. Three input beams, the \emph{sources}, carry binary signals. Each need has a different \emph{valid} source that tracks its hidden value; the third source is unrelated to both. The network must learn these assignments without being supplied their labels. We later change one assignment to test whether the same network can revise a preference that was previously useful. The current answer and the learned relevance of a source are stored in different physical variables.

Each source connects to both answer memories through separate links, giving six links in total. A link contains a cavity and a second resonator, the \emph{reader}, which couples the source to the answer memory. An excitation reservoir shared by both modes holds the \emph{relevance record}. Its population shifts both resonances, as excitons shift optical modes in polariton microcavities~\cite{stepanov2019,kalinin2019}. The cavity response to a fixed-frequency probe biases source selection. The reader response to the source field sets how strongly that source changes the answer. One population therefore controls both intake and influence through two physical readouts.

The experimenter supplies the timing, source-selection rule and feedback. The needs act in turn. A weak probe measures the acting need's links, and a fixed rule selects a source with probability favouring stronger cavity responses. The selected source then acts on both answer memories through their readers. An external scorer checks the acting need's answer. A common outcome pulse drives the \emph{writer}. Each record update is weighted by a local \emph{trace}, a temporary measure of recent link activity. The writer's contribution is the link's \emph{credit}. A separate correction drive resets the acting answer memory without writing relevance records. For binary answers, the action and its score already determine this correction. We set the feedback threshold so that merely retaining the corrected answer gives a mean outcome signal close to zero (Methods).

A useful source must first have a chance to correct a wrong answer. Initially, unassisted coupling is too weak to reverse the stored answer. On half of all readings, \emph{assisted reading} tunes the selected reader of the acting need to resonance, regardless of source identity. This lets a source revise a wrong answer and earn credit before its relevance record forms (Supplementary Sections~S1.3 and S3.2). The assistance supplies access without identifying useful sources or prescribing target records.

Learning therefore requires a record to remain both stored and usable. Further reinforcement changes the physical response through which it is read. It can shift the record past a resonance, so that the population persists while its influence weakens. We derive conditions for useful readout and use them to design local writing rules. Their tests cover learning, continued operation and reassignment. The physical relation between storage, readout and writing thus determines whether experience can continue to guide the choice and use of information.

\begin{table}[!t]
\centering
\small
\caption{\textbf{Glossary.} Model components and terms. Symbols follow Methods.}
\label{tab:glossary}
\begin{tabular}{@{}p{0.27\linewidth}p{0.69\linewidth}@{}}
\toprule
Term & Meaning \\
\midrule
Need (E, I) & One of two binary questions, answered in turn. The \emph{acting} need is the one answered this cycle. \\
Source & Input beam carrying a binary signal in its phase. A need's \emph{valid} source tracks its hidden value; their pairing is the \emph{assignment}. \\
Answer memory $z_j$ & Parametric oscillator whose phase, $0$ or $\pi$, stores need $j$'s answer. \\
Link & Cavity, reader and reservoir connecting a source to an answer memory; six links in total. \\
Cavity $b_{jk}$ & Continuously driven resonator whose probed response biases source selection. \\
Reader $c_{jk}$ & Resonator coupling a source to memory; its gain sets the source's influence. \\
Relevance record $x_{jk}$ & Shared reservoir population shifting both resonances. It is the learned state; \emph{formed} when $x_{jk}>x_U$. \\
Sensing, protection, revision & Selecting valid sources, preventing reversal by non-valid sources, and permitting revision by valid sources. Their common allowed range is the \emph{functional window}. \\
Trace $\epsilon_{jk}$ & Local activity retained until feedback: cavity light (co-presence), reader light (admission), or reader work on memory (use). \\
Writer $\Pi_{jk}$ & Common outcome pulse weighted by local traces. Self-limiting writing weakens positive updates as records form. \\
Correction drive $u_j$ & Resets the acting answer after scoring without writing relevance records. \\
Assisted reading & On half of readings, the acting need's selected reader is tuned to resonance, allowing revision before record formation. \\
Adverse-outcome susceptibility $\mu_{jk}$ & Signed local state raised by adverse outcomes; its positive part strengthens negative writing. \\
\bottomrule
\end{tabular}
\end{table}

\section*{One population stores the record and sets both readouts}

A continuous holding drive sustains optical bistability in each link. When the cavity field relaxes much faster than the reservoir, filling and decay reduce to
\begin{equation}
 \dot x/\gamma=-x+A/D(x),\qquad D(x)=1+(\delta-x)^2,
 \label{eq:red}
\end{equation}
where $x$ is the normalised population, $\gamma$ its decay rate, $A$ the drive power and $\delta$ the detuning. At $\delta=3$ and holding power $A=A_0=4$, the stable populations are $x_L=2-\sqrt2$ and $x_H=2+\sqrt2$, separated by $x_U=2$~\cite{baas2004}. A record is \emph{formed} when $x>x_U$; the drive sustains it after writing ends.

The settled sensing response is proportional to $1/D(x)$. The same population sets reader coupling to the answer memory, whose stable states without drives are $z_j=\pm1$. Source $k$ carries $y_k=\pm3$. For a rapidly relaxing reader,
\begin{equation}
 \left.\dot z_j\right|_{\rm readers}\simeq
 \sum_k I_kK(x_{jk})(y_k-z_j),
 \label{eq:reader}
\end{equation}
where $I_k$ marks the selected source and $K(x)=g_0/\{1+\mathrm i(x-x_H)/w\}$, with width $w$ and peak gain $g_0$. Its real part, $g(x)=\operatorname{Re}K(x)$, pulls the answer towards the source. A weak pull leaves the answer in its original well; sufficient gain removes the opposing stable state and permits revision (Methods). Sensing peaks at $x=3$, whereas the reader is designed to peak at $x_H$. Growth beyond both peaks weakens selection and influence while leaving the record formed.

\section*{Action outcomes establish endogenous information control}

We first test learning at reader width $w=1$. After a non-guess action, the outcome pulse adds
\begin{equation}
 \Pi_{jk}=\Lambda\sigma\epsilon_{jk}H(x_{jk})\,Q,
 \qquad Q=s-\beta_{j^*}-\vartheta,
 \qquad H(x)=\frac{x}{x+x_*}.
 \label{eq:writer}
\end{equation}
to the population dynamics. Here $s$ is one for a correct answer and zero otherwise, $j^*$ is the acting need, and $\beta_{j^*}$ estimates chance performance from designated fair guesses, which occupy one action in five and cause no writing. The margin is $\vartheta=0.2$, the writing gain is $\Lambda$, and $\sigma$ marks the two-unit pulse. The local trace $\epsilon_{jk}$ decays when activity ceases; $H$, with $x_*=1$, keeps the writer from driving populations negative. This follows three-factor learning, in which delayed reward acts on recently active connections~\cite{yagishita2014,brzosko2015,izhikevich2007,gerstner2018}.

\begin{figure}[!t]
\centering
\includegraphics[width=\linewidth]{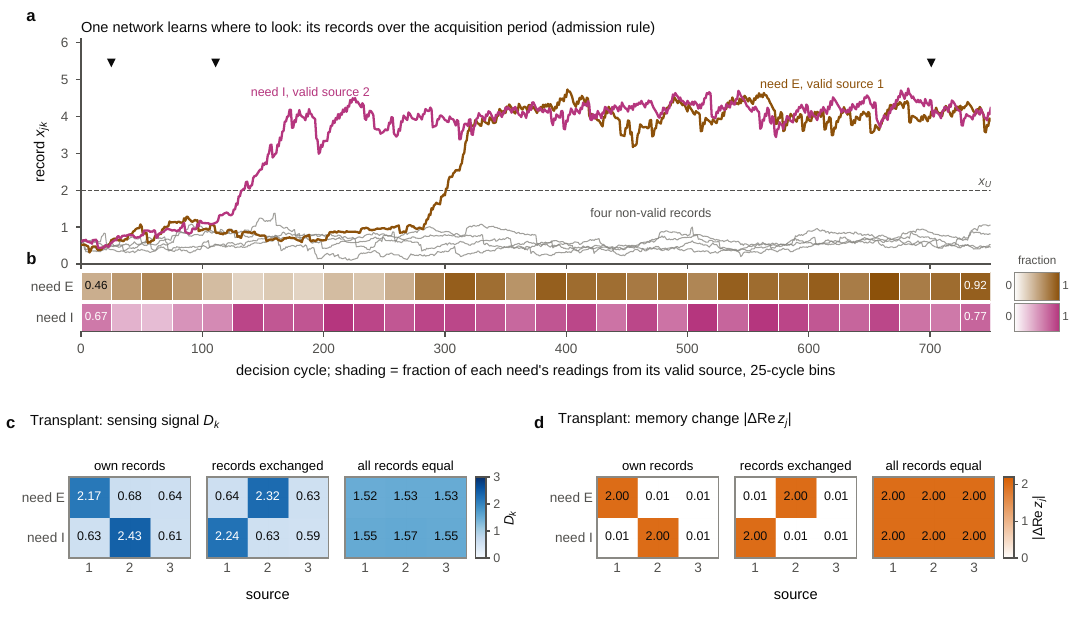}
\caption{\textbf{Learned populations transfer both source choice and memory influence.}
\textbf{a}, Records during acquisition (cycles 1--750) in one of the $24$ admission networks of Fig.~\ref{fig:cohort}, whose traces are driven by reader intensity. Colours: valid records of need E (brown) and need I (magenta), and four non-valid records (grey). Dashed line, $x_U$; triangles, the three cycles shown in Supplementary Fig.~S2.
\textbf{b}, Fraction of each need's readings taken from its valid source, in 25-cycle bins (white 0, full colour 1); the first and last bins are labelled.
\textbf{c},\textbf{d}, Transplant in the first eight-pair admission test, with criteria fixed before the runs. Each pair contains two networks trained on exchanged assignments. One network per pair is tested with its own records, its partner's six records (records exchanged) and all six records set to their mean (all records equal). Memories, traces and fast fields are matched; only the reservoir populations are exchanged.
\textbf{c}, Sensing signal $D_k$, the cavity intensity of source $k$, measured two time units after the named need's probe is switched on, with readers off.
\textbf{d}, Memory change: the largest change of $\operatorname{Re}z_j$, the answer memory of need $j$, over the two signs of the source value after a ten-unit reading with the probe off. With equal records, every source changes the memory (Supplementary Section~S7.3).
Full network equations with noise (Methods; reference parameters, $w=1$). Panels \textbf{a},\textbf{b} show a single realisation, recomputed with logging that leaves the dynamics unchanged (Supplementary Section~S9.2). Panels \textbf{c},\textbf{d} show means over the eight pairs; ranges are given in the text. No error bars.}
\label{fig:learn}
\end{figure}

We compare traces of cavity light (\emph{co-presence}), reader intensity (\emph{admission}) and reader-induced memory motion (\emph{use}). Admission and use accumulate only while the link's need acts. In the illustrated admission network, both valid records form while the four non-valid records remain unformed (Fig.~\ref{fig:learn}a,b). In unassisted cycles, the probability of selecting need E's valid source rises from $0.40$ before formation to $0.88$ afterwards. The formed records also permit unassisted revision (Supplementary Fig.~S2).

To test whether the populations carry both learned functions, we exchange only the six populations between networks trained on exchanged assignments, matching all other preparations (Fig.~\ref{fig:learn}c,d). Source preference and memory influence follow the transplanted records in all eight initial pairs, with criteria fixed before the runs. The preferred source's sensing signal exceeds its strongest competitor by $2.35$--$4.0$; a source valid under the transplanted pattern produces memory displacement $2.00$, against at most $0.019$ for the others. Equalising all six populations removes selectivity. In the larger cohort, the same test passed in $21/24$ admission pairs and in $15/16$ use pairs with both networks formed (Supplementary Section~S7.3).

\begin{figure}[!t]
\centering
\includegraphics[width=\linewidth]{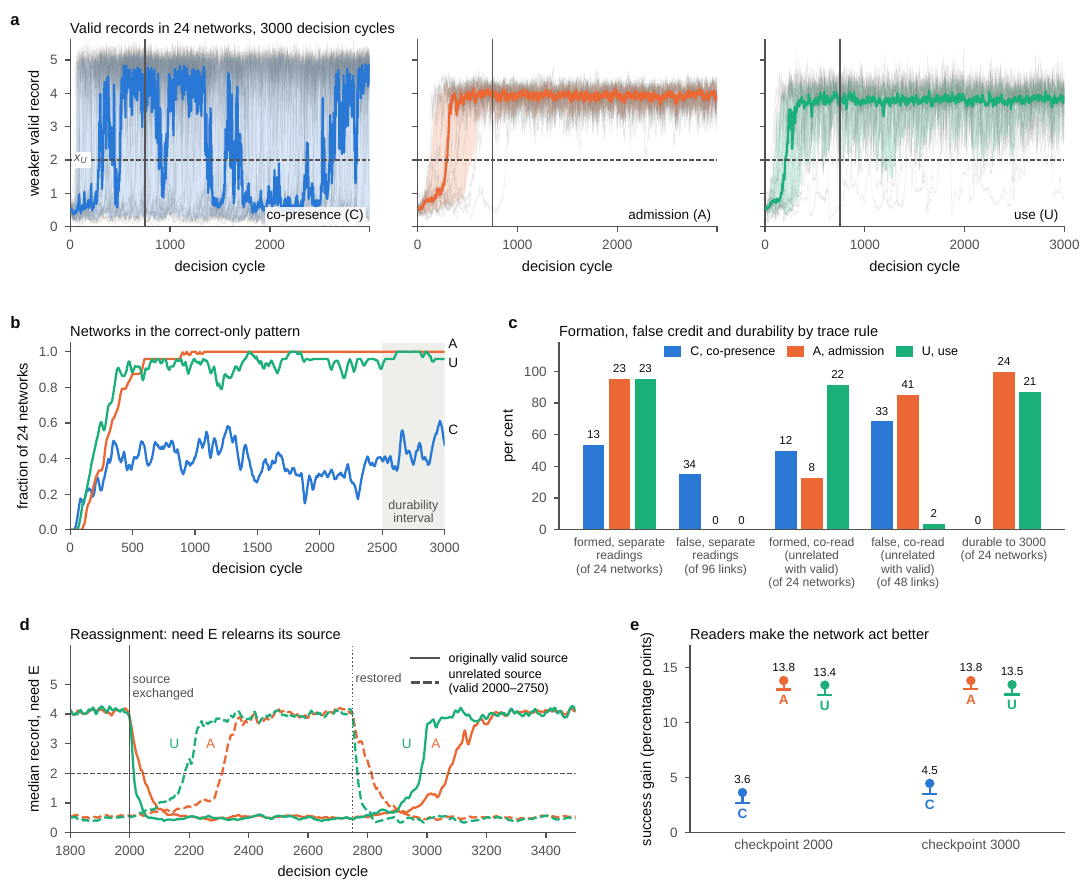}
\caption{\textbf{Traces tied to reading support formation, retention and replacement.}
Twenty-four networks per trace rule, full noisy equations, $w=1$, $\Delta t=0.002$ (Methods). C, co-presence; A, admission; U, use.
\textbf{a}, Weaker valid record in each network (thin lines), median (thick line) and 10--90\% band. Dashed line, $x_U=2$; vertical line, acquisition deadline at cycle 750.
\textbf{b}, Fraction in the correct-only pattern (both valid records formed, all non-valid records unformed), 25-cycle moving average. Shading marks cycles 2501--3000.
\textbf{c}, Separate readings: networks with both valid records formed at cycle 750 and persistent false records among 96 non-valid links. Co-reading shows the same quantities with the unrelated source added to each valid reading (48 co-read links). Durability requires the correct-only pattern for at least 400 of cycles 2501--3000 and no persistent false record in cycles 1--3000. Bars show percentages; labels show counts. The registered co-reading false-record test requires both valid records to form in at least 18/24 networks; only use qualifies.
\textbf{d}, Need E's valid and unrelated sources are exchanged at cycle 2000 and restored at 2751; runs end at 3500. For admission and use, curves show medians over 24 networks (9-cycle moving average) of the original valid-source record (solid) and unrelated-source record (dashed).
\textbf{e}, Paired 200-cycle continuations from cycles 2000 and 3000 compare connected and disconnected readers. Both retain writing and correction; assistance is off. Symbols show the mean gain in correct actions, including designated guesses, in percentage points. Bars are one-sided 95\% lower bounds from network bootstraps. Without readers, retaining a corrected answer can still succeed.
Networks are the sampling units. Curves in \textbf{a}, \textbf{b} and \textbf{d} include all networks and have no error bars. Some counts change under step refinement; aggregate verdicts do not (Supplementary Section~S9.1).}
\label{fig:cohort}
\end{figure}

Across $24$ networks per trace rule, admission and use each form both valid records in $23$, with no persistent false records among $96$ non-valid links at the reference integration step. Co-presence forms both valid records in $13$ and leaves $34$ persistent false records (Fig.~\ref{fig:cohort}). Score shuffling preserves the mean score but causes false records under every rule, establishing the importance of outcome--reading associations. Disconnecting readers with assistance off reduces correct actions by $13$--$14$ percentage points. Both rules form newly valid records after reassignment without changed-link labels. Admission retains records more consistently under separate readings and is used below. Under the flip protocol, only use passes the joint criteria when an unrelated source is co-read with the valid one (Supplementary Sections~S7.2--S7.4).

\section*{Storage stability is not functional stability}
At the reference reader width $w=1$, learned records persist and improve actions, but settle near $x=4$, beyond both readout peaks. Positive writing continues after formation, while the restoring flow, set by slow reservoir decay, is too weak to hold the record near $x_H$ (Supplementary Sections~S1.1 and S6).

\begin{figure}[!t]
\centering
\includegraphics[width=\linewidth]{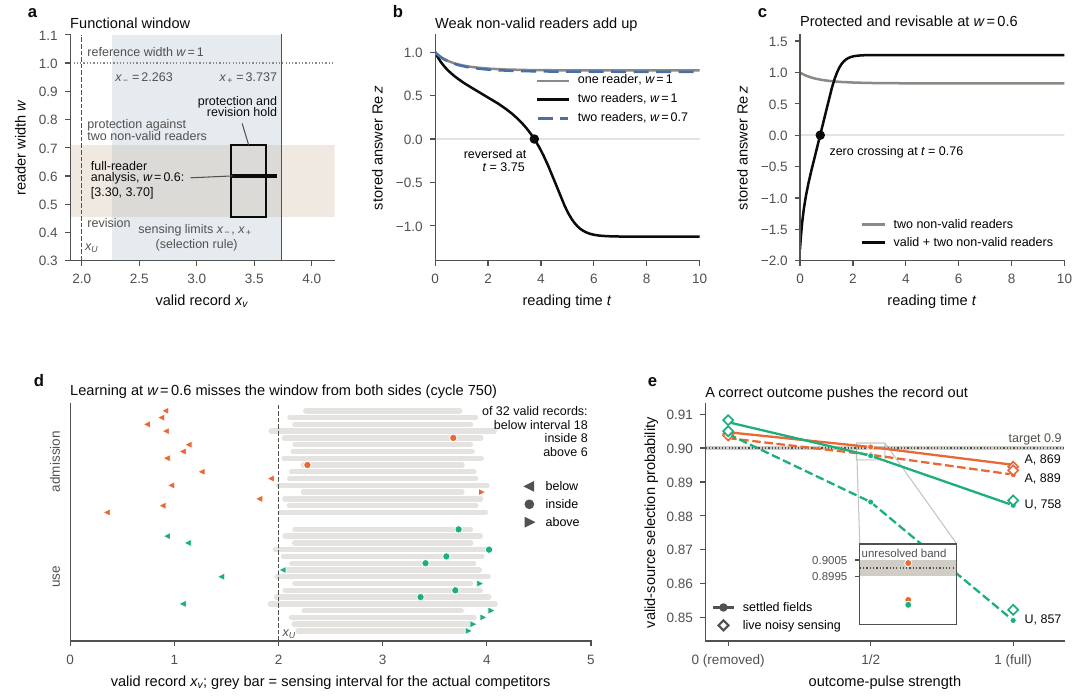}
\caption{\textbf{A stored record need not be a functional record.}
\textbf{a}, Functional window for fixed records, with non-valid records in $[0.50,0.75]$. The selection rule, Eq.~(\ref{eq:selection}), guarantees sensing for valid records $x_v$ between $x_-=2.263$ and $x_+=3.737$ for settled cavity fields (grey-blue band). Bounds from the reduced equations (Methods) require protection against two opposing unformed readers and valid revision within one reading. These conditions give the width range certified by the bounds (beige band, $0.4551\le w<0.7101$, valid records in $[3.30,3.60]$, box); the band marks what the bounds certify, not every functional configuration. The bar marks the range proved with the full reader equations at $w=0.6$, with valid records in $[3.30,3.70]$.
\textbf{b}, From $z(0)=+1$, the record of each active reader at $x_L$ and each source value $-3$: one reader at $w=1$ and two readers at $w=0.7$ keep the answer; two at $w=1$ reverse it at $t=3.75$.
\textbf{c}, At $w=0.6$, two non-valid readers with records $0.75$ and source values $-3$ leave an initially positive answer unchanged in sign. Against these readers, a valid reader with record $3.60$ and source value $+3$ revises an answer starting at $z(0)=-1.8$, the largest displacement the bounds cover, crossing zero at $t=0.76$. Panels \textbf{b},\textbf{c} use reference parameters (Methods), with readers starting at zero.
\textbf{d}, Learning at $w=0.6$ (eight networks per trace rule, two needs each; Supplementary Section~S7.7) at the acquisition deadline, cycle 750. For each need, the grey interval gives the valid-record values that ensure selection probability at least $0.9$ against its two non-valid records; the symbol shows its valid record (orange, admission; green, use).
\textbf{e}, Four rewarded cycles from the histories shown in \textbf{d}, each replayed from the same state and random inputs with the common six-link outcome pulse at full, half or zero strength. Filled symbols, settled-field selection probability for the end records; open symbols, live noisy sensing, tested at full and zero strength only. Labels give the trace rule (A, admission; U, use) and event cycle. The grey band, $0.8995$--$0.9005$, marks the undecided sensing range (Methods); inset, half strength enlarged.
Panel \textbf{a} is analytical. Panels \textbf{b},\textbf{c} integrate the full reader and memory equations without noise; \textbf{d},\textbf{e} show single realisations of the full network equations with noise. No error bars.}
\label{fig:window}
\end{figure}

We call a record configuration functional when three conditions hold: each need selects its valid source with probability at least $p_*=0.9$ (sensing); unassisted non-valid readings cannot reverse a stored answer (protection); and an unassisted valid source can revise the answer within one reading (revision). Their common range is the \emph{functional window} (Fig.~\ref{fig:window}a--c).

Sensing restricts a valid record on both sides of the probe resonance, $x=3$: too far from resonance, it no longer outshines its competitors enough. The competing populations and supplied selection rule determine the lower and upper sensing limits, $x_-$ and $x_+$. Protection depends on the summed gain of non-valid readers. At $w=1$, two readers attached to unformed records can together reverse an answer, although either is harmless alone. We require protection against both simultaneously, a margin stronger than the standard protocol's single non-valid reading (Methods).

Narrowing the reader leaves its peak gain unchanged but suppresses far-detuned, unformed records more strongly than valid records near $x_H$. Valid records occupy a range of populations, however, and excessive narrowing can leave those near its edges too weak to revise the answer. Reduced-equation bounds identify a sufficient common width range, $0.4551\le w<0.7101$, which excludes $w=1$; they do not rule out narrower readers (Methods; Fig.~\ref{fig:window}a). We use $w=0.6$. At this width, the full reader and memory equations establish all three functions for fixed valid records in $[3.30,3.70]$, with non-valid records in $[0.50,0.75]$ and the initial reader and memory states specified in Methods. Related bounds cover slowly moving and noisy records under additional conditions.

These results certify readout for specified record configurations. They do not establish that learning reaches or preserves those configurations. The writer must bring records across the formation barrier $x_U$ and keep them within the functional window.

\section*{Success can push a record out of its functional window}
With the original writer at $w=0.6$ and $\Lambda=5$, records formed but rarely settled inside the functional window (Fig.~\ref{fig:window}d). Continuations from learned states lost sensing mostly through its upper boundary (Methods).

We tested whether reinforcement causes this loss by replaying four rewarded cycles from saved states with matched random inputs and full, half or zero outcome-pulse strength (Fig.~\ref{fig:window}e). The pulse writes all six links, changing valid and competing records together. In every event, the full pulse pushed a valid record already beyond both readout peaks farther from resonance. Competing populations differed by at most $0.017$ between full-pulse and pulse-off replays. Its settled-field selection probability fell to $0.849$--$0.895$; removing the pulse kept it above $0.9$. The record remained formed in every replay. Reinforcement can therefore increase the stored population while weakening source selection. The intervention establishes causation in these events, which were selected after inspection; it does not estimate their frequency (Supplementary Section~S7.8).

\section*{A self-limiting writer keeps learned records useful}
\begin{figure}[!tp]
\centering
\includegraphics[width=0.97\linewidth]{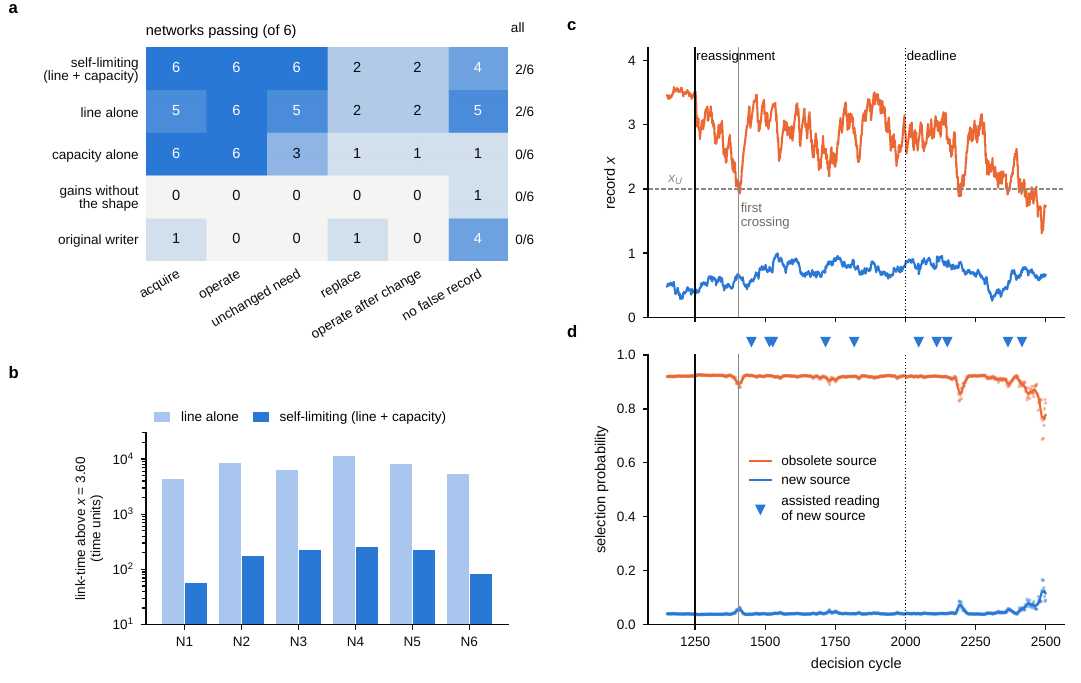}\\[2pt]
\includegraphics[width=0.97\linewidth]{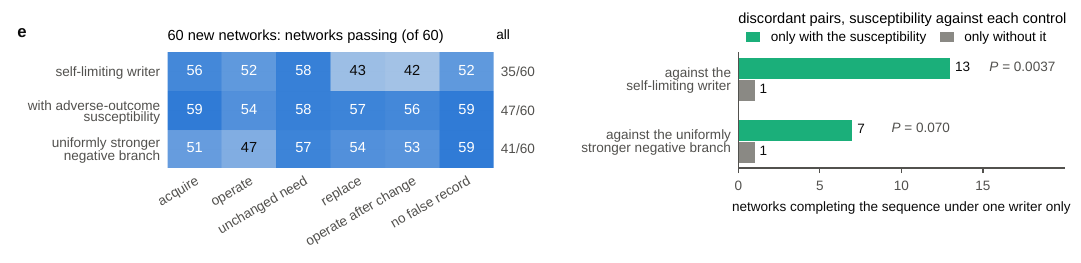}
\caption{\textbf{Preserving readout does not ensure timely replacement.}
\textbf{a}, Counts among networks N1--N6, one per reassignment class, passing each criterion fixed before the runs: functional acquisition by cycle 750, operation to cycle 1250, the unchanged need, replacement by cycle 2000, operation after the change to cycle 2500 and no persistent false record. Right, networks passing all six. Rows show five writer laws: the self-limiting writer (absorption line + capacity ceiling), Eq.~(\ref{eq:self-limiting}); the line alone; the capacity alone at $\Lambda_+=8.75$; the line's gains without its shape; and the original writer ($\Lambda=5$).
\textbf{b}, Summed excursion time above $x=3.60$ during cycles 0--2499 in networks N1--N6, comparing the line alone with the self-limiting writer (logarithmic scale, time units). An excursion begins at an upward crossing of $3.60$ and ends when the record falls below $3.59$.
\textbf{c},\textbf{d}, Replacement in network N6 under the self-limiting writer. Panel \textbf{c} shows end-of-cycle records of the changed need's obsolete (orange) and new (blue) sources. Panel \textbf{d} shows their selection probabilities at that need's decisions (dots; lines, 9-decision moving averages). Triangles above \textbf{d} mark assisted readings of the new source followed by outcome writing. Vertical lines mark reassignment (cycle 1250), the obsolete record's first crossing below $x_U$ and the replacement deadline (cycle 2000). Supplementary Section~S7.11 tabulates all six histories.
\textbf{e}, Counts among 60 new networks, ten per reassignment, with random streams matched across writers (Methods). The writers are the self-limiting rule, its extension with the adverse-outcome susceptibility, and a uniformly stronger negative branch calibrated to the extension's added negative writing on the development histories. Right, completion of the whole sequence; counts and tests are given in the text.
Full network equations with noise, run in the original simulation code (Methods) at $\Delta t=0.0005$, $w=0.6$, with the admission trace. Panels \textbf{a} and \textbf{e} use criteria fixed before the runs; \textbf{b}--\textbf{d} were analysed after the runs. Panels \textbf{c},\textbf{d} show single realisations. No error bars.}
\label{fig:s3val}
\end{figure}

The writer must promote formation and then limit further growth. An approximate mean-drift calculation using measured admission statistics finds no constant gain that meets both requirements (Methods). We therefore shape only positive writing by the link's own population, without a validity label or sensing signal:
\begin{equation}
 \Pi_{jk}=\sigma\epsilon_{jk}H(x_{jk})
 \bigl\{\Lambda_+f_w(x_{jk})[Q]_+-\Lambda_-[-Q]_+\bigr\},
 \qquad f_w(x)=a_w(x)\,\kappa_{\rm cap}(x),
 \label{eq:self-limiting}
\end{equation}
Here $[Q]_+=\max(Q,0)$, and $\Lambda_+$ and $\Lambda_-$ set positive and negative writing. The remaining symbols retain their definitions in Eq.~(\ref{eq:writer}). The absorption line $a_w(x)=1/\{1+[(\delta_w-x)/w_w]^2\}$ weakens positive writing as the record forms. The capacity $\kappa_{\rm cap}(x)=\min\{1,\max[0,(x_c-x)/(x_c-x_a)]\}$ removes it at a ceiling $x_c$ inside the functional window. With $f_w\equiv1$ and $\Lambda_\pm=\Lambda$, the original writer is recovered.

We use $w_w=1$, $\delta_w=1.25$, $\Lambda_+=8.75$, $\Lambda_-=5$, $x_a=3.30$ and $x_c=3.60$. The negative branch remains unchanged. The absorption line follows from the cavity response to a separate write tone. Its resonance lies below the formation barrier, so absorption is stronger near the barrier than near the formed state. Trace and outcome control the write power as in the original rule. The capacity remains phenomenological and requires an additional filling mechanism (Methods). Positive writing alone cannot cross its ceiling; confinement under passive flow and noise needs separate conditions. The readout physics thus determines where reinforcement should stop.

We tested the rule in six networks, N1--N6, one per reassignment, without resetting between learning, operation and adaptation. All six acquired functional records by the deadline, operated until reassignment and preserved the unchanged need, satisfying the first three registered criteria (Fig.~\ref{fig:s3val}a). The same gains without either population factor acquired functional records in none ($P=0.002$, Fisher's exact test added after the runs); the original writer succeeded in one (Methods).

The remaining criteria tested replacement, operation after reassignment and absence of persistent false records. Adding the capacity to the absorption line reduced total excursion time above $3.60$ about fortyfold and eliminated excursions above $3.70$ in these histories (Fig.~\ref{fig:s3val}b). Nevertheless, only two networks completed all six criteria. Functional replacement requires readout through the new source, a stronger condition than record formation. The other four histories exposed a further obstacle.

\section*{An obsolete record can block its own replacement}

An obsolete record can restrict the readings that could establish its replacement. In N5 and N6, the two most severe replacement failures, its source remained the most probable choice at every decision of the changed need from the record's first dip below $x_U$ until the replacement deadline. Its mean selection probability was $0.88$ and $0.92$, compared with $0.065$ and $0.04$ for the new source. These measurements were made after the runs (Fig.~\ref{fig:s3val}c,d for N6; Supplementary Section~S7.11).

Crossing below $x_U$ does not remove the record's sensing advantage over a weaker competitor. Its population can also recover when passive filling and restoration outweigh negative writing. To test its causal role, we restarted N5 and N6 at cycle $1626$ and reset only the obsolete population to $x_L$, matching all other initial states and random streams. The new source then had the stronger sensing signal at nearly every later decision of the changed need. Both networks passed replacement, while the unchanged need continued to pass its sensing tests (Supplementary Section~S7.12). The obsolete record therefore obstructed replacement. Because the reset changes both readouts, it does not separate source-selection effects from reader coupling.

\section*{Adverse outcomes release obsolete records}

To release an obsolete record through learning, the writer must respond to local activity and outcomes without a signal identifying which source has changed. We add a signed, decaying state to each link, its \emph{adverse-outcome susceptibility}. Adverse outcome signals raise this state and favourable signals lower it, with updates weighted by the link's eligibility trace. It also relaxes towards zero. Only its positive part strengthens negative writing. The aim is to target links whose recent activity repeatedly accompanies unsuccessful actions; accumulation depends on trace-weighted outcomes, not error frequency alone. Positive writing is unchanged, and the reservoir population still controls both readouts (Methods).

In the six development networks, susceptibility became positive on obsolete links after reassignment. It remained non-positive on formed valid links during the tested operation periods, leaving their negative writing unchanged. One parameter was tuned using the saved histories of N5 and N6 (Supplementary Section~S7.12).

We tested the extension on $60$ new networks, ten per reassignment, with random streams matched across writers and the cohort, endpoint and tests fixed before the runs (Supplementary Section~S7.13). Success required all six criteria of Fig.~\ref{fig:s3val}a to pass in the same history. The extension completed the sequence in $47$ networks against $35$ for the self-limiting writer: $13$ succeeded only with the extension and one only without it (Fig.~\ref{fig:s3val}e). This is a paired improvement of $20$ percentage points (two-sided exact paired test, Holm-adjusted $P=0.0037$). The largest gains were in replacement, operation after the change and absence of persistent false records. Local sensitivity to adverse outcomes therefore increases the probability of completing the learning sequence under the tested conditions.

A uniform control with stronger negative writing, calibrated to the extension's added writing on saved development histories, completed $41$ networks. The extension's advantage, seven discordant networks against one, did not reach significance (Holm-adjusted $P=0.070$). How much of the benefit requires history dependence, beyond the added negative writing, remains open.

To connect local record dynamics to these outcomes, we constructed a reduced stochastic model. Its two measured inputs, the trace area of a reading and the probability that a reading reverses the answer, were calibrated retrospectively from the self-limiting writer's histories in the sixty-network comparison. Holding that calibration fixed across writers, the model predicted $30.7$, $44.4$ and $37.5$ complete sequences per $60$ networks for the self-limiting writer, extension and uniform control. No parameter was fitted to these aggregate counts. The predicted gains of the extension were $13.7$ and $7.0$, compared with $12$ and $6$ observed. Replacement was systematically underpredicted, chiefly through late formation after reassignment. This retrospective comparison supports transfer of the calibrated dynamics across writing rules at this operating point (Methods; Supplementary Section~S7.14 and Fig.~S4).

\section*{Implications for physical learning}

Learning which source to consult changes the conditions for further learning. A source that has helped before can become more likely to be sampled and more able to change the answer. The relevance records therefore shape the evidence from which they themselves are updated. Here, two resonant readouts of one persistent population give this feedback loop a physical basis. Adaptation requires the loop to remain open to evidence that could revise an earlier choice.

Stable storage alone cannot ensure this. A record can remain in its formed state while its readouts become too weak to support useful behaviour. Bounded plasticity and history-dependent updates are familiar in synapses and devices~\cite{grossberg1980,fusi2005,fusiabbott2007,bcm1982,vanrossum2000,loeffler2023,prezioso2016,rahman2023}. The constraint here comes from how the record is used. Our sufficient bounds identify a range where source selection, protection against misleading signals and revision by useful signals coexist. Leaving that range can destroy function without erasing the record. This differs from synaptic saturation~\cite{fusiabbott2007} or destruction of an attractor by training~\cite{essex2026,eisenmann2023}. The readout consequently supplies a physical criterion for where reinforcement should stop.

The obsolete-record result adds a second requirement: learned preferences must remain open to revision when usefulness changes. A previously successful source can retain its selection advantage and restrict access to its replacement. Local responses to adverse outcomes improve replacement in the model, although the benefit of history dependence beyond stronger negative writing remains unresolved. The retrospectively calibrated stochastic reduction connects local response statistics to changes in sequence completion across writers. No parameter was fitted to aggregate completion rates. The comparison therefore tests whether calibrated local dynamics transfer between writing rules at this operating point. The model underpredicts the formation of new records late after reassignment.

A similar readout constraint can arise wherever a persistent state shifts a resonance read at a fixed frequency. Applying the learning mechanism additionally requires local activity traces, outcome-dependent writing and one state controlling both information intake and influence. Microrings provide precedents for persistent resonance shifts~\cite{foradori2026} and state-dependent routing coupled to later writing~\cite{lugnan2025}. Combining these ingredients in one functioning learning link remains an experimental question.

A candidate polariton link would use two optical modes sharing one exciton reservoir. Reservoir-induced shifts~\cite{schmidt2019}, depletion of non-radiative reservoirs~\cite{kurdyubov2021} and electrically controlled exciton ionisation~\cite{suchomel2017} have been demonstrated separately. The first experiment can be small: prepare several reservoir populations and measure both the sensing response and the reader's influence on memory (Supplementary Section~S8). This tests whether both readouts reproducibly follow the same stored state before attempting learning. A full implementation still needs physical selection, outcome-gated writing and mechanisms for the trace, capacity limit and susceptibility. The present demonstration is theoretical and numerical; scaling with network size and noise remains open. Measuring these response curves would identify a candidate operating range to test during learning.

\section*{Methods}

\subsection*{Field--reservoir element and network}

Optical cavities coupled to excitation reservoirs support neuromorphic computing~\cite{opala2019,ballarini2020}.

Absorbed light fills a reservoir whose population shifts the driven cavity resonance. The cavity amplitude $a$ and population $n$ obey
\begin{equation}
 \dot a=[-\kappa+\mathrm i(\Delta-g_Rn)]a+f,
 \qquad \dot n=s_{\rm cap}|a|^2-\gamma n,
 \label{eq:elem}
\end{equation}
where $\kappa$ and $\gamma$ are field and reservoir decay rates, $f$ is the drive amplitude and $\Delta$ its detuning above the bare resonance. The coefficients $g_R$ and $s_{\rm cap}$ set the shift per excitation and capture rate; $s_{\rm cap}$ is distinct from the score $s$. For $\kappa\gg\gamma$, eliminating the field with $x=g_Rn/\kappa$ and $\delta=\Delta/\kappa$ gives Eq.~(\ref{eq:red}). The holding power gives low and high stable record states (Supplementary Section~S1.1).

The network simulations retain both optical fields. For the link from source $k$ to need $j$, $b_{jk}$ is the cavity field, $c_{jk}$ the reader field, $x_{jk}$ the reservoir population and $z_j$ the answer memory. In memory-response units,
\begin{align}
 \dot b_{jk}&=\kappa\{[-1+\mathrm i(\delta-x_{jk})]b_{jk}+\tilde f_{jk}\},
 &\tilde f_{jk}&=\sqrt{A_0}+\mathrm i\sqrt B(P_j+I_k),
 \label{eq:network-field}\\
 \dot x_{jk}&=\gamma(|b_{jk}|^2-x_{jk}+\chi|c_{jk}|^2)+\Pi_{jk},
 \label{eq:network-record}\\
 \dot c_{jk}&=-[\kappa_r+\mathrm i\Delta_{r,jk}(t)]c_{jk}
 -\mathrm i vI_k(z_j-y_k),
 \label{eq:network-reader}\\
 \dot z_j&=\left[\frac{1-G}{2}-|z_j|^2\right]z_j
 +\frac{1+G}{2}z_j^*
 -\mathrm i\sum_kvI_kc_{jk}+u_j.
 \label{eq:network-memory}
\end{align}
The reader decays at rate $\kappa_r$ and couples reciprocally to memory with strength $v$~\cite{takata2015reader}, giving $g_0=v^2/\kappa_r$. The prescribed source is $y_k=3\zeta_k$, $\zeta_k=\pm1$. The parameter $G$ damps the imaginary memory coordinate. Without reader or correction drives, the stable states are $z_j=\pm1$, the normal form of a degenerate optical parametric oscillator above threshold~\cite{wang2013opo}.

The holding drive remains on. Binary gates $P_j$ and $I_k$ control the acting need's probe and the selected source's weak tone and readers; the readers couple that source to both memories. The weak tones, each of power $B$, are mutually in phase and in quadrature with holding. At $B=0.2$, their maximum combined drive power is $4.8$, below the quasistatic switching threshold $5.089$. Readers have no holding drive. Ordinary detuning is $\Delta_{r,jk}=\kappa_r(x_{jk}-x_H)/w$; assistance sets it to zero only on the acting need's selected link. The coefficient $\chi$ sets reader-light absorption. Only $\Pi_{jk}$ receives outcomes. After scoring, $u_{j^*}=3\zeta_{j^*}$ corrects the acting answer without writing relevance records, where $\zeta_{j^*}$ is that need's current hidden value.

Resonance-shifting feedback occurs in polariton-reservoir, free-carrier and photothermal cavities~\cite{stepanov2019,wouters2007,hamerly2015,konthasinghe2017}. Its reduction depends on relaxation rates~\cite{kalinin2019}. A polariton implementation requires a shared reservoir shifting both readouts and a direct polariton self-interaction shift small enough to preserve field linearity within the readout error budget. Supplementary Table~S2 gives reference parameters; Section~S8 gives implementation requirements. Record formation at the slower-field setting has not established the full learning sequence (Supplementary Section~S7.5).

\subsection*{Selection, parameters and decision cycle}

During sensing, the acting need's cavity intensities give
\begin{equation}
 D_k=\sum_jP_j|b_{jk}|^2 ,
 \label{eq:sensing}
\end{equation}
where $D_k$ is distinct from the denominator $D(x)$ in Eq.~(\ref{eq:red}). Source $k$ is selected with probability
\begin{equation}
 p_k=(1-\epsilon_x)\frac{\exp(\beta_sD_k/D_\Sigma)}
 {\sum_l\exp(\beta_sD_l/D_\Sigma)}
 +\frac{\epsilon_x}{N_K},
 \qquad D_\Sigma=\sum_lD_l.
 \label{eq:selection}
\end{equation}
The first term favours stronger relative responses, with $\beta_s$ setting the preference strength. The second provides uniform exploration. We use $N_K=3$, $\beta_s=8$ and $\epsilon_x=0.1$, giving a maximum selection probability of about $0.933$, even with perfect sensing contrast. After selection, an independent draw provides assisted reading with probability $0.5$. This allows a source to revise an answer before its record has formed.

Each run starts with $x_{jk}=x_L$ and $b_{jk}=c_{jk}=z_j=\epsilon_{jk}=0$. The network evolves for $100$ time units under the holding drive alone, with noise. The task then starts with newly drawn hidden values and $\beta_j=0.5$. Supplementary Table~S2 lists the reference parameters.

The needs alternate in $22$-unit cycles (Fig.~\ref{fig:cycle}c). Each cycle contains two units of sensing, ten of reading and a two-unit delay before scoring. The acting need's probe stays on until scoring. Correction then acts for four units; after non-guess actions, the outcome pulse writes during the first two. Four quiet units complete the cycle.

With probability $0.2$, the scored action is a designated fair guess. Its score updates the chance baseline,
$\beta_{j^*}\leftarrow(1-\alpha)\beta_{j^*}+\alpha s$,
with $\alpha=0.02$, and no outcome pulse follows. Otherwise, the action is the sign of $\operatorname{Re}z_{j^*}$, with a fair coin used when $|\operatorname{Re}z_{j^*}|<0.2$. This tie-break remains a non-guess action and triggers writing.

Hidden values flip with probability $0.1$ at the end of each reading and each cycle. A copied valid reading therefore succeeds with probability $0.9$. Retaining the corrected answer from the same need's previous action succeeds with probability $0.705$, after four opportunities to flip. Each source also carries its own background binary value, drawn by a fair coin when the task starts and flipped independently like the hidden values; a source valid for no need reports it. After reassignment, the obsolete source therefore reports its background value and the new valid source its need's hidden value. Supplementary Section~S1.3 gives the full protocol and its redraw and co-reading variants.

\subsection*{Writer and traces}

The writer factor $H(x)=x/(x+x_*)$, $x_*=1$, vanishes at zero. At $\beta_j=0.5$, correct and incorrect answers give $Q=+0.3$ and $-0.7$. Mean outcomes are $-0.2$ for chance success and $+0.005$ for retaining the previous correction. Actual credit also depends on the outcome's association with the trace (Eq.~(\ref{eq:trace-credit-criterion})).

The traces retain activity with lifetime $T_\epsilon$ and suppress inputs below $\epsilon_0$. Phase-change resistance drift has been proposed to implement physical traces~\cite{demirag2021}. Co-presence measures cavity light above holding, admission reader intensity, and use the reader's contribution along the memory's motion:
\begin{align}
 T_\epsilon\dot\epsilon^{\rm c}_{jk}
 &=\bigl[|b_{jk}|^2-A_0/D(x_{jk})-\epsilon_0\bigr]_+
 -\epsilon^{\rm c}_{jk},
 \label{eq:elig}\\
 T_\epsilon\dot\epsilon^{\rm a}_{jk}
 &=\bigl[P_jS_a|c_{jk}|^2-\epsilon_0\bigr]_+
 -\epsilon^{\rm a}_{jk},
 \label{eq:adm}\\
 T_\epsilon\dot\epsilon^{\rm u}_{jk}
 &=\bigl[P_jS_u\operatorname{Re}
 \{\dot z_j^{\,*}(-\mathrm i vI_kc_{jk})\}-\epsilon_0\bigr]_+
 -\epsilon^{\rm u}_{jk}.
 \label{eq:use}
\end{align}
Here $[\cdot]_+=\max(\cdot,0)$, and $S_a,S_u$ set activity scales. Each simulation uses one rule with parameters in Supplementary Table~S2. The cutoff $\epsilon_0$ differs from exploration $\epsilon_x$. The gate $P_j$ restricts admission and use to the acting need and switches off during feedback, leaving residual traces to write.

Writing $z_j=q+\mathrm ip$, the intrinsic memory is a gradient flow in $\mathcal V=(q^2+p^2)^2/4-q^2/2+Gp^2/2$. During reading, correction is off and the forces $r_{jk}=-\mathrm ivI_kc_{jk}$ satisfy
\begin{equation}
 \sum_k\operatorname{Re}\bigl(\dot z_j^{\,*}r_{jk}\bigr)
 =\dot{\mathcal V}+|\dot z_j|^2 .
 \label{eq:use-work}
\end{equation}
These are work rates in an effective potential, not physical energy-transfer rates. Gating, thresholding and rectification make credit allocation approximate. Use distinguishes confirming and reversing readings and can suppress unrelated co-read credit (Supplementary Section~S2), but is task-dependent: redrawing hidden values every cycle produces persistent false records on $17/48$ co-read links (Supplementary Section~S7.2). When a need stops acting, its admission and use traces decay and their remaining negative writing is finite (Supplementary Section~S4.1).

\subsection*{Learned relevance, tests and statistics}

Formed records are $S_{jk}=\mathbf1\{x_{jk}>x_U\}$; $V_{jk}$ separately specifies source validity. Learned relevance requires outcomes to write populations without access to $V$, satisfying five criteria: (D1) outcomes produce $S=V$ from preparation, unlike signal co-occurrence alone; (D2) learned records increase source-selection probability; (D3) they increase influence at fixed input and memory state; (D4) changing the acting need changes preferences without rewriting $S$, whereas reassignment changes only its affected entries; (D5) one local rule works across tested assignments. Source relabelling preserves the homogeneous equations at fixed size (Supplementary Section~S4.2).

R1--R5 used criteria fixed before runs; G1 replayed and rescored them (Supplementary Section~S9.2). G2 compared traces on $24$ new networks per rule with matched inputs. G3 extended those histories to $3000$ cycles and added interventions.

Seeded networks are the sampling units. Network proportions use exact binomial intervals; link-rate intervals resample networks. Trace-rule counts use exact paired tests for matched streams and Fisher's test otherwise. Median waiting times use Kaplan--Meier estimates with right censoring. Reader-withdrawal effects use one-sided paired-network bootstrap intervals. Transplants change only populations under common preparation of memories, traces and fields. Criteria, thresholds and schedules are in Supplementary Sections~S7.1--S7.4; writer-study designs follow below.

\subsection*{Functional readout bounds}

If the valid sensing signal exceeds each competitor by a factor of at least $R_I>1$, Eq.~(\ref{eq:selection}) gives
\begin{equation}
 p_v\ge\frac{\epsilon_x}{N_K}+
 \frac{1-\epsilon_x}{1+(N_K-1)e^{-\beta_s(R_I-1)/(R_I+N_K-1)}}.
 \label{eq:functional-sensing}
\end{equation}
At the reference settings, $R_I\ge3.93$ ensures $p_v\ge0.9$. For settled fields and non-valid records in $[0.50,0.75]$, this gives $[x_-,x_+]=[2.263,3.737]$. The limits depend on selection settings, target probability and competitors (Supplementary Section~S3).

For protection, write $q=\operatorname{Re}z_j$ and require a margin $m>0$ from $q=0$. Within the state bounds of Supplementary Section~S3.3, an answer initially satisfying $|q|\ge m$ keeps its sign against sources $|y_k|\le v_0$ with summed non-valid gain at most $\overline G_d$ if
\begin{equation}
 m-m^3-\overline G_d(v_0+m)-\eta_r>0,
 \label{eq:functional-protection}
\end{equation}
Here $\eta_r$ bounds the total residual in the reduced equation, including reader transients and complex-response corrections. The imaginary part of $K(x)$ vanishes at resonance and acts mainly on the strongly damped imaginary memory coordinate.

At $w=1$, two readers at $x_L$ have combined gain $0.156$, above the reversal threshold $0.108$. We require protection against two simultaneous non-valid readers, although standard readings present at most one; the stronger test covers co-reading. Revision requires the valid reader to overcome restoring forces and allowed opposition within one reading. For valid records in $[3.30,3.60]$, non-valid records in $[0.50,0.75]$ and $\eta_r=0.05$, the sufficient common range is $0.4551\le w<0.7101$. These guarantees concern unassisted readings; assistance lies outside them.

At $w=0.6$, the full-reader analysis removes the assumed residual bound, includes switch-on transients and proves all three functions for fixed valid records in $[3.30,3.70]$ with the stated non-valid records and reset states. It also covers record paths staying within $0.01$ of reference values, with valid references in $[3.30,3.60]$, under the stated initial-state and competitor bounds (Supplementary Sections~S3.3--S3.4). Sensing additionally requires bounded cavity-field errors, including direct field noise. The analysis assigns no probability to remaining in these ranges and gives no guarantee from arbitrary dynamical states.

\subsection*{Finite pulses and the learning target}

The credit delivered by a finite pulse depends on the trace throughout that pulse. For $x>0$, the coordinate $\Psi(x)=x+x_*\ln(x/x_*)$ removes the factor $H$ from the original writer. The writer's contribution to the change in $\Psi$ is exactly $J=\mathcal A(s-\nu)$, where $\mathcal A=\Lambda\int\sigma\epsilon\,\mathrm dt$ is the trace area over the pulse and $\nu=\beta_{j^*}+\vartheta$ is constant during it. Passive flow and noise contribute separately. For fixed $\nu$ and $\mathbb E\mathcal A>0$,
\begin{equation}
 \mathbb E J>0\quad\Longleftrightarrow\quad
 \frac{\mathbb E[\mathcal A s]}{\mathbb E\mathcal A}>\nu .
 \label{eq:trace-credit-criterion}
\end{equation}
Credit therefore depends on success weighted by local activity. The overall fraction of successful actions does not determine it. Even at equal populations, two links can receive different credit from the same outcome pulse because their traces differ (Supplementary Section~S5.2).

The learning target is the probability that, from the acquisition deadline through operation, the records remain in the functional window and every allowed readout test succeeds from the evolving state (Supplementary Section~S5.3). The simulations check acquisition by a deadline and the readings sampled during operation; they do not certify continuous survival. With the original writer at $w=0.6$ and $\Lambda=5$, all tested continuations from learned states lost sensing within $151$ cycles, mostly through the upper boundary of the sensing interval (Supplementary Section~S7.9).

\subsection*{Matched outcome-pulse intervention}

The pulse-replay test (FRCP1) asks whether a rewarded pulse can cause a loss of sensing. We replayed four positive pulses that preceded such a loss in the learning histories behind Fig.~\ref{fig:window}d (FPL1). Each replay starts from the complete saved network state with the original random streams. Only the writer term on all six links changes: it is multiplied by $M=1,\tfrac12,0$, giving a full, half or absent pulse.

The full-pulse replay reproduces the original event, and all three versions agree before the pulse. Differences afterwards can therefore be attributed to the changed writing. The events were selected after inspection; the multipliers and thresholds for classifying sensing were fixed before replay. The test establishes causation in these selected events without estimating how often the mechanism occurs. Supplementary Section~S7.8 gives the event identifiers, settled-field and live sensing measurements, and the band in which sensing is classified as unresolved.

\subsection*{Construction of the self-limiting writer}

A write tone of power $A_w$ and detuning $\delta_w$ drives the holding mode. At fixed $x$, fields add linearly and beat at $\kappa|\delta-\delta_w|\gg\gamma$. After cavity relaxation, with the envelope and reservoir slow over a beat period,
\[
 \langle|b|^2\rangle\simeq
 \frac{A_0}{D(x)}+\frac{A_w}{1+(\delta_w-x)^2}.
\]
Power controlled by the trace and positive outcome supplies $a_w(x)$ with $w_w=1$. This control and $H(x)$ remain proposed device operations. Representative formation pulses require roughly ten times the holding power without saturating the cavity or reservoir (Supplementary Section~S6).

The capacity is one below $x_a$, falls linearly to zero at $x_c$ and vanishes above it. Positive writing alone cannot reach $x_c$ from below in finite time. At $x_c$, settled cavity filling remains below decay even with both weak tones; reader absorption and transients must fit within the remaining margin for deterministic confinement. Noise still produces excursions: six-network maxima were $3.63$--$3.66$ against $x_c=3.60$. This phenomenological capacity requires a separate filling channel that tracks $x$ rapidly. Saturating the shared reservoir would also change holding and bistability (Supplementary Section~S6).

Writer parameters came from an approximate mean-flow calculation combining passive restoration with measured original-writer event rates, trace areas and outcomes. It seeks positive drift through formation, a high equilibrium below the functional limit and negative obsolete-record drift under chance-level outcomes. Under admission statistics, formation requires $\Lambda\gtrsim7.4$, while confinement below $x_+$ requires $\Lambda\lesssim5.5$ (or $4.9$ for the proved reader window). No constant gain satisfies both in this approximation. Use statistics overlap narrowly and depend on conditioning traces on outcomes. At $\Lambda=5$, admission's slightly negative formation drift is consistent with fluctuation-assisted formation. Because event statistics may change with the writer, this is a design guide tested by controls (Supplementary Section~S6 and Fig.~S3).

\subsection*{Development study}

FML1 used a separate implementation at $\Delta t=0.002$, eight networks per writer/trace combination and $w=0.6$. Streams were matched across writers on the schedule below. Every cycle was checked for sensing and applicable unassisted protection/revision. Controls separated gain changes, absorption, capacity and branch choice. Under admission, uniformly stronger writing and the self-limiting gains without population factors each formed records but left no network functional at acquisition; shaping both branches erased no obsolete record. At $\Delta t=0.002$, formation was faster than in the original code. At $0.0005$, sampled formation-time distributions were consistent, but individual-trajectory agreement remains unestablished. Development and validation results are not pooled (Supplementary Section~S7.10).

\subsection*{Six-network validation}

We ran networks N1--N6 (study S3VAL) in the original code at $\Delta t=0.0005$ and $w=0.6$, using the admission trace. Each network represented one of the six combinations of changed need and new valid source. Runs followed four stages without reset: acquisition to cycle $750$, operation until reassignment at $1250$, adaptation to $2000$ and further operation to $2500$. We compared the five writer laws in Fig.~\ref{fig:s3val}a. The line with capacity and the line alone shared random streams.

Six criteria were fixed before the runs: functional acquisition; operation before reassignment; preservation of the unchanged need; replacement by cycle $2000$; operation after that deadline; and absence of persistent false records. Functional acquisition required formed records with successful sensing, protection and revision. Replacement required sensing and revision through the new source, together with protection against every tested pattern of non-valid readings containing the obsolete source. Supplementary Section~S7.11 defines the pass and fail thresholds and the undecided bands between them. Two separately written scorers agreed on all $180$ criterion labels.

The comparisons fixed before the runs concerned completion of the whole sequence. The acquisition comparison of six successes against none was added afterwards. It used Fisher's exact test because the self-limiting writer and the control with the same gains but neither population factor did not share random streams. Selection probabilities, counts of new-source readings and excursion times were also analysed after the runs. Supplementary Section~S7.11 reports these analyses.

\subsection*{Adverse-outcome susceptibility and sixty-network comparison}

Each link has a signed susceptibility $\mu_{jk}$, initialised to zero at preparation. During feedback, it accumulates $\sigma\epsilon_{jk}$ with weight $+1$ for $Q<0$ and $-r_\mu$ for $Q>0$. It also relaxes towards zero with lifetime $T_\mu$. The negative gain in Eq.~(\ref{eq:self-limiting}) becomes $\Lambda_-(1+\eta_\mu[\mu_{jk}]_+)$, while positive writing is unchanged. The extension requires no additional external signal and reduces to the self-limiting writer wherever $\mu_{jk}\le0$ (Supplementary Section~S7.12).

We used $r_\mu=1/3$, $T_\mu=2200$ and $\eta_\mu=1.3$. The gain $\eta_\mu$ was selected using the saved histories of N5 and N6 during the development study (RLR1), and $T_\mu$ was chosen with knowledge of the reassignment schedule. All three parameters were fixed before the first simulation with the extension and were not changed afterwards.

The sixty-network comparison (VAL1) tested the self-limiting writer, its susceptibility extension and a uniform control on $60$ new networks, ten per reassignment. All started from the unformed preparation in the original code at $\Delta t=0.0005$ and followed the schedule above. Random streams were matched across writers within each network. The uniform control retained the positive branch and multiplied the negative gain by approximately $1.228$, giving $\Lambda_{-,U}=6.142$. This multiplier was calibrated before the comparison by evaluating the extension along the six saved self-limiting-writer histories. Summed over all links and histories, its added negative writing in $\Psi$ was $0.228$ times the baseline negative writing. This calibration matches writing on the saved paths; the new trajectories can receive different totals.

The cohort, endpoint, two primary comparisons, tests and decision rule were fixed before the runs. The endpoint required all six criteria to pass in the same history. Two-sided exact McNemar tests, which compare the extension with each control using only the networks that passed under one writer and failed under the other, were the primary tests. A Holm adjustment, which multiplies the smaller of the two $P$-values by two, controlled the familywise error rate at $0.05$ across the two comparisons. Superiority required statistical significance and a positive paired difference. Numerically unresolved histories counted as failures. The comparison between the uniform control and the self-limiting writer was secondary. Two separately written scorers agreed on all $1{,}080$ criterion labels. The failed development expectation that initial formation times would remain within $\pm10\%$ of those under the self-limiting writer was retained as a secondary diagnostic (Supplementary Section~S7.13).

\subsection*{Reduced stochastic model of the learning sequence}

The reduction was constructed after the sixty-network comparison. Once per cycle, it advances the six records with their pulse trace areas, the binary answer memories, the baselines, the hidden and background values and, for the extension, the susceptibilities; the records alone are not assumed to be Markovian. Selection follows Eq.~(\ref{eq:selection}) with settled sensing signals. An assisted reading sets the answer memory to the source value; an unassisted opposing reading does so with a measured probability. Each link then receives the writer's contribution in the coordinate $\Psi$, with its state-dependent factors evaluated at the start of the cycle, together with passive filling at the cycle-averaged drive power and Gaussian record noise of the prescribed strength. The two measured inputs, trace areas and reversal probabilities as functions of assistance, the read link's record and whether the source value agrees with the stored answer, come from the self-limiting writer's VAL1 histories and were held fixed across writers (Supplementary Section~S7.14).

For each writer, $10{,}000$ replicas started from the unformed preparation and followed the $2500$-cycle VAL1 schedule; rates are expected counts per $60$ networks. Applied to the $180$ VAL1 histories, a record-space scorer reproduces $1{,}073$ of the $1{,}080$ criterion labels and the registered replacement and complete-sequence counts. No parameter was fitted to aggregate replacement or complete-sequence counts. The comparison is retrospective and does not change the registered VAL1 tests or their conclusions.

\subsection*{Noise and numerical resolution}

We integrate the deterministic dynamics with fourth-order Runge--Kutta steps. After each step, independent Gaussian increments are added to each record and to the real and imaginary parts of each cavity field. Negative populations are set to zero. Readers, memories, traces and susceptibility states receive no direct noise (Supplementary Sections~S1.5 and S7.12).

The trace-rule comparison and its extension (G2, G3) use $\Delta t=0.002$, with matched-noise checks at $0.001$ and, where needed, $0.0005$. All aggregate test verdicts were unchanged across the tested steps, although individual counts and some quantities measured over short trajectory segments depended on the step size (Supplementary Section~S9).

The six-network and sixty-network writer studies use $\Delta t=0.0005$. A separate check (RES1) repeated the six development networks N1--N6 under the three VAL1 writers at $\Delta t=0.00025$, with noise increments coupled between resolutions. All $18$ complete-sequence outcomes and $108$ criterion labels were unchanged. This tests sensitivity to one halving of the step; it does not establish a convergence order (Supplementary Section~S9.1).

\section*{Data availability}
The source data for the figures, the per-cycle numerical histories and the scored criterion labels are available in Zenodo at \url{https://doi.org/10.5281/zenodo.23775077} (ref.~\citenum{ERTdeposit}). Within-cycle observer records are regenerated from the deposited code and seeds. Supplementary Information, Section~S9.3 lists the contents of the deposit.

\section*{Code availability}
The simulation, analysis and scoring code, including the reduced-model implementation and calibration tables, is available in the same Zenodo record (\url{https://doi.org/10.5281/zenodo.23775077}; ref.~\citenum{ERTdeposit}).

\section*{Acknowledgements}
N.G.B.~acknowledges support from the HORIZON EIC-2022-PATHFINDERCHALLENGES-01 HEISINGBERG Project 101114978, the EPSRC UK Multidisciplinary Centre for Neuromorphic Computing (grant UKRI982) and the Weizmann--UK Make Connection Grant 142568.
The author thanks Anthropic for access to Claude, a large language model.
Claude was used to edit the manuscript and the Supplementary Information, and Claude Code was used to help write and run the simulation and analysis code.
The author checked all code, results and text and takes full responsibility for the content.

\section*{Author contributions}
N.G.B.~conceived the study, developed the model and its analysis, designed and carried out the numerical simulations, analysed the results and wrote the manuscript.

\section*{Competing interests}
The author declares no competing interests.

\section*{Additional information}
\textbf{Supplementary information} is available at \url{https://www.damtp.cam.ac.uk/user/ngb23/publications/endogenous_si.pdf}.
Correspondence and requests for materials should be addressed to N.G.B.

\clearpage
\renewcommand{\bibfont}{\small}


\begin{thebibliography}{10}
\def\ERTgobble, {}
\providecommand{\ERTdoi}[2]{\href{https://doi.org/#1}{\ERTgobble#2}}
\providecommand{\ERTurl}[2]{\href{#1}{\ERTgobble#2}}
\providecommand{\ERTnolink}[2]{\ERTgobble#2}
\expandafter\ifx\csname url\endcsname\relax
  \def\url#1{\texttt{#1}}\fi
\providecommand{\bibinfo}[2]{#2}


\bibitem{mackintosh1975}
\bibinfo{author}{Mackintosh, N.~J.}
\newblock \bibinfo{title}{A theory of attention: variations in the
  associability of stimuli with reinforcement}.
\newblock \ERTdoi {10.1037/h0076778} {, \emph{\bibinfo{journal}{Psychol. Rev.}}
  \textbf{\bibinfo{volume}{82}}, \bibinfo{pages}{276--298}
  (\bibinfo{year}{1975})}.

\bibitem{pearcehall1980}
\bibinfo{author}{Pearce, J.~M.} \& \bibinfo{author}{Hall, G.}
\newblock \bibinfo{title}{A model for {Pavlovian} learning: variations in the
  effectiveness of conditioned but not of unconditioned stimuli}.
\newblock \ERTdoi {10.1037/0033-295X.87.6.532} {,
  \emph{\bibinfo{journal}{Psychol. Rev.}} \textbf{\bibinfo{volume}{87}},
  \bibinfo{pages}{532--552} (\bibinfo{year}{1980})}.

\bibitem{gottlieb2013}
\bibinfo{author}{Gottlieb, J.}, \bibinfo{author}{Oudeyer, P.-Y.},
  \bibinfo{author}{Lopes, M.} \& \bibinfo{author}{Baranes, A.}
\newblock \bibinfo{title}{Information-seeking, curiosity, and attention:
  computational and neural mechanisms}.
\newblock \ERTdoi {10.1016/j.tics.2013.09.001} {,
  \emph{\bibinfo{journal}{Trends Cogn. Sci.}} \textbf{\bibinfo{volume}{17}},
  \bibinfo{pages}{585--593} (\bibinfo{year}{2013})}.

\bibitem{keim2019}
\bibinfo{author}{Keim, N.~C.}, \bibinfo{author}{Paulsen, J.~D.},
  \bibinfo{author}{Zeravcic, Z.}, \bibinfo{author}{Sastry, S.} \&
  \bibinfo{author}{Nagel, S.~R.}
\newblock \bibinfo{title}{Memory formation in matter}.
\newblock \ERTdoi {10.1103/RevModPhys.91.035002} {,
  \emph{\bibinfo{journal}{Rev. Mod. Phys.}} \textbf{\bibinfo{volume}{91}},
  \bibinfo{pages}{035002} (\bibinfo{year}{2019})}.

\bibitem{kaspar2021}
\bibinfo{author}{Kaspar, C.}, \bibinfo{author}{Ravoo, B.~J.},
  \bibinfo{author}{van~der Wiel, W.~G.}, \bibinfo{author}{Wegner, S.~V.} \&
  \bibinfo{author}{Pernice, W. H.~P.}
\newblock \bibinfo{title}{The rise of intelligent matter}.
\newblock \ERTdoi {10.1038/s41586-021-03453-y} {,
  \emph{\bibinfo{journal}{Nature}} \textbf{\bibinfo{volume}{594}},
  \bibinfo{pages}{345--355} (\bibinfo{year}{2021})}.

\bibitem{bhattacharyya2022}
\bibinfo{author}{Bhattacharyya, K.}, \bibinfo{author}{Zwicker, D.} \&
  \bibinfo{author}{Alim, K.}
\newblock \bibinfo{title}{Memory formation in adaptive networks}.
\newblock \ERTdoi {10.1103/PhysRevLett.129.028101} {,
  \emph{\bibinfo{journal}{Phys. Rev. Lett.}} \textbf{\bibinfo{volume}{129}},
  \bibinfo{pages}{028101} (\bibinfo{year}{2022})}.

\bibitem{stern2021}
\bibinfo{author}{Stern, M.}, \bibinfo{author}{Hexner, D.},
  \bibinfo{author}{Rocks, J.~W.} \& \bibinfo{author}{Liu, A.~J.}
\newblock \bibinfo{title}{Supervised learning in physical networks: from
  machine learning to learning machines}.
\newblock \ERTdoi {10.1103/PhysRevX.11.021045} {, \emph{\bibinfo{journal}{Phys.
  Rev. X}} \textbf{\bibinfo{volume}{11}}, \bibinfo{pages}{021045}
  (\bibinfo{year}{2021})}.

\bibitem{stern2023}
\bibinfo{author}{Stern, M.} \& \bibinfo{author}{Murugan, A.}
\newblock \bibinfo{title}{Learning without neurons in physical systems}.
\newblock \ERTdoi {10.1146/annurev-conmatphys-040821-113439} {,
  \emph{\bibinfo{journal}{Annu. Rev. Condens. Matter Phys.}}
  \textbf{\bibinfo{volume}{14}}, \bibinfo{pages}{417--441}
  (\bibinfo{year}{2023})}.

\bibitem{dillavou2022}
\bibinfo{author}{Dillavou, S.}, \bibinfo{author}{Stern, M.},
  \bibinfo{author}{Liu, A.~J.} \& \bibinfo{author}{Durian, D.~J.}
\newblock \bibinfo{title}{Demonstration of decentralized physics-driven
  learning}.
\newblock \ERTdoi {10.1103/PhysRevApplied.18.014040} {,
  \emph{\bibinfo{journal}{Phys. Rev. Applied}} \textbf{\bibinfo{volume}{18}},
  \bibinfo{pages}{014040} (\bibinfo{year}{2022})}.

\bibitem{scellier2017}
\bibinfo{author}{Scellier, B.} \& \bibinfo{author}{Bengio, Y.}
\newblock \bibinfo{title}{Equilibrium propagation: bridging the gap between
  energy-based models and backpropagation}.
\newblock \ERTdoi {10.3389/fncom.2017.00024} {, \emph{\bibinfo{journal}{Front.
  Comput. Neurosci.}} \textbf{\bibinfo{volume}{11}}, \bibinfo{pages}{24}
  (\bibinfo{year}{2017})}.

\bibitem{lopezpastor2023}
\bibinfo{author}{L{\'o}pez-Pastor, V.} \& \bibinfo{author}{Marquardt, F.}
\newblock \bibinfo{title}{Self-learning machines based on {Hamiltonian} echo
  backpropagation}.
\newblock \ERTdoi {10.1103/PhysRevX.13.031020} {, \emph{\bibinfo{journal}{Phys.
  Rev. X}} \textbf{\bibinfo{volume}{13}}, \bibinfo{pages}{031020}
  (\bibinfo{year}{2023})}.

\bibitem{wright2022}
\bibinfo{author}{Wright, L.~G.} \emph{et~al.}
\newblock \bibinfo{title}{Deep physical neural networks trained with
  backpropagation}.
\newblock \ERTdoi {10.1038/s41586-021-04223-6} {,
  \emph{\bibinfo{journal}{Nature}} \textbf{\bibinfo{volume}{601}},
  \bibinfo{pages}{549--555} (\bibinfo{year}{2022})}.

\bibitem{momeni2023}
\bibinfo{author}{Momeni, A.}, \bibinfo{author}{Rahmani, B.},
  \bibinfo{author}{Mall{\'e}jac, M.}, \bibinfo{author}{del Hougne, P.} \&
  \bibinfo{author}{Fleury, R.}
\newblock \bibinfo{title}{Backpropagation-free training of deep physical neural
  networks}.
\newblock \ERTdoi {10.1126/science.adi8474} {,
  \emph{\bibinfo{journal}{Science}} \textbf{\bibinfo{volume}{382}},
  \bibinfo{pages}{1297--1303} (\bibinfo{year}{2023})}.

\bibitem{ibrahim2026}
\bibinfo{author}{Ibrahim, M.}
\newblock \bibinfo{title}{Sequential learning and catastrophic forgetting in
  differentiable resistor networks}.
\newblock \ERTdoi {10.1103/584w-2mxq} {, \emph{\bibinfo{journal}{Phys. Rev. E}}
  \textbf{\bibinfo{volume}{113}}, \bibinfo{pages}{065303}
  (\bibinfo{year}{2026})}.

\bibitem{du2026}
\bibinfo{author}{Du, Y.}, \bibinfo{author}{van Mastrigt, R.},
  \bibinfo{author}{Veenstra, J.} \& \bibinfo{author}{Coulais, C.}
\newblock \bibinfo{title}{Metamaterials that learn to change shape}.
\newblock \ERTdoi {10.1038/s41567-026-03226-2} {, \emph{\bibinfo{journal}{Nat.
  Phys.}} \textbf{\bibinfo{volume}{22}}, \bibinfo{pages}{784--790}
  (\bibinfo{year}{2026})}.

\bibitem{dillavou2024}
\bibinfo{author}{Dillavou, S.} \emph{et~al.}
\newblock \bibinfo{title}{Machine learning without a processor: emergent
  learning in a nonlinear analog network}.
\newblock \ERTdoi {10.1073/pnas.2319718121} {, \emph{\bibinfo{journal}{Proc.
  Natl Acad. Sci. USA}} \textbf{\bibinfo{volume}{121}},
  \bibinfo{pages}{e2319718121} (\bibinfo{year}{2024})}.

\bibitem{feldmann2019}
\bibinfo{author}{Feldmann, J.}, \bibinfo{author}{Youngblood, N.},
  \bibinfo{author}{Wright, C.~D.}, \bibinfo{author}{Bhaskaran, H.} \&
  \bibinfo{author}{Pernice, W. H.~P.}
\newblock \bibinfo{title}{All-optical spiking neurosynaptic networks with
  self-learning capabilities}.
\newblock \ERTdoi {10.1038/s41586-019-1157-8} {,
  \emph{\bibinfo{journal}{Nature}} \textbf{\bibinfo{volume}{569}},
  \bibinfo{pages}{208--214} (\bibinfo{year}{2019})}.

\bibitem{loeffler2023}
\bibinfo{author}{Loeffler, A.} \emph{et~al.}
\newblock \bibinfo{title}{Neuromorphic learning, working memory, and
  metaplasticity in nanowire networks}.
\newblock \ERTdoi {10.1126/sciadv.adg3289} {, \emph{\bibinfo{journal}{Sci.
  Adv.}} \textbf{\bibinfo{volume}{9}}, \bibinfo{pages}{eadg3289}
  (\bibinfo{year}{2023})}.

\bibitem{dillavou2025}
\bibinfo{author}{Dillavou, S.} \& \bibinfo{author}{Mishra, S.}
\newblock \bibinfo{title}{Physical reinforcement learning}.
\newblock \bibinfo{howpublished}{Preprint at
  \url{https://arxiv.org/abs/2511.17789}} (\bibinfo{year}{2025}).

\bibitem{lugnan2025}
\bibinfo{author}{Lugnan, A.} \emph{et~al.}
\newblock \bibinfo{title}{Emergent self-adaptation in an integrated photonic
  neural network for backpropagation-free learning}.
\newblock \ERTdoi {10.1002/advs.202404920} {, \emph{\bibinfo{journal}{Adv.
  Sci.}} \textbf{\bibinfo{volume}{12}}, \bibinfo{pages}{2404920}
  (\bibinfo{year}{2025})}.

\bibitem{kanno2020}
\bibinfo{author}{Kanno, K.}, \bibinfo{author}{Naruse, M.} \&
  \bibinfo{author}{Uchida, A.}
\newblock \bibinfo{title}{Adaptive model selection in photonic reservoir
  computing by reinforcement learning}.
\newblock \ERTdoi {10.1038/s41598-020-66441-8} {, \emph{\bibinfo{journal}{Sci.
  Rep.}} \textbf{\bibinfo{volume}{10}}, \bibinfo{pages}{10062}
  (\bibinfo{year}{2020})}.

\bibitem{kanno2022}
\bibinfo{author}{Kanno, K.} \& \bibinfo{author}{Uchida, A.}
\newblock \bibinfo{title}{Photonic reinforcement learning based on
  optoelectronic reservoir computing}.
\newblock \ERTdoi {10.1038/s41598-022-07404-z} {, \emph{\bibinfo{journal}{Sci.
  Rep.}} \textbf{\bibinfo{volume}{12}}, \bibinfo{pages}{3720}
  (\bibinfo{year}{2022})}.

\bibitem{stepanov2019}
\bibinfo{author}{Stepanov, P.} \emph{et~al.}
\newblock \bibinfo{title}{Dispersion relation of the collective excitations in
  a resonantly driven polariton fluid}.
\newblock \ERTdoi {10.1038/s41467-019-11886-3} {, \emph{\bibinfo{journal}{Nat.
  Commun.}} \textbf{\bibinfo{volume}{10}}, \bibinfo{pages}{3869}
  (\bibinfo{year}{2019})}.

\bibitem{kalinin2019}
\bibinfo{author}{Kalinin, K.~P.} \& \bibinfo{author}{Berloff, N.~G.}
\newblock \bibinfo{title}{Polaritonic network as a paradigm for dynamics of
  coupled oscillators}.
\newblock \ERTdoi {10.1103/PhysRevB.100.245306} {,
  \emph{\bibinfo{journal}{Phys. Rev. B}} \textbf{\bibinfo{volume}{100}},
  \bibinfo{pages}{245306} (\bibinfo{year}{2019})}.

\bibitem{baas2004}
\bibinfo{author}{Baas, A.}, \bibinfo{author}{Karr, J.~{\relax Ph}.},
  \bibinfo{author}{Eleuch, H.} \& \bibinfo{author}{Giacobino, E.}
\newblock \bibinfo{title}{Optical bistability in semiconductor microcavities}.
\newblock \ERTdoi {10.1103/PhysRevA.69.023809} {, \emph{\bibinfo{journal}{Phys.
  Rev. A}} \textbf{\bibinfo{volume}{69}}, \bibinfo{pages}{023809}
  (\bibinfo{year}{2004})}.

\bibitem{yagishita2014}
\bibinfo{author}{Yagishita, S.} \emph{et~al.}
\newblock \bibinfo{title}{A critical time window for dopamine actions on the
  structural plasticity of dendritic spines}.
\newblock \ERTdoi {10.1126/science.1255514} {,
  \emph{\bibinfo{journal}{Science}} \textbf{\bibinfo{volume}{345}},
  \bibinfo{pages}{1616--1620} (\bibinfo{year}{2014})}.

\bibitem{brzosko2015}
\bibinfo{author}{Brzosko, Z.}, \bibinfo{author}{Schultz, W.} \&
  \bibinfo{author}{Paulsen, O.}
\newblock \bibinfo{title}{Retroactive modulation of spike timing-dependent
  plasticity by dopamine}.
\newblock \ERTdoi {10.7554/eLife.09685} {, \emph{\bibinfo{journal}{eLife}}
  \textbf{\bibinfo{volume}{4}}, \bibinfo{pages}{e09685}
  (\bibinfo{year}{2015})}.

\bibitem{izhikevich2007}
\bibinfo{author}{Izhikevich, E.~M.}
\newblock \bibinfo{title}{Solving the distal reward problem through linkage of
  {STDP} and dopamine signaling}.
\newblock \ERTdoi {10.1093/cercor/bhl152} {, \emph{\bibinfo{journal}{Cereb.
  Cortex}} \textbf{\bibinfo{volume}{17}}, \bibinfo{pages}{2443--2452}
  (\bibinfo{year}{2007})}.

\bibitem{gerstner2018}
\bibinfo{author}{Gerstner, W.}, \bibinfo{author}{Lehmann, M.},
  \bibinfo{author}{Liakoni, V.}, \bibinfo{author}{Corneil, D.} \&
  \bibinfo{author}{Brea, J.}
\newblock \bibinfo{title}{Eligibility traces and plasticity on behavioral time
  scales: experimental support of {neoHebbian} three-factor learning rules}.
\newblock \ERTdoi {10.3389/fncir.2018.00053} {, \emph{\bibinfo{journal}{Front.
  Neural Circuits}} \textbf{\bibinfo{volume}{12}}, \bibinfo{pages}{53}
  (\bibinfo{year}{2018})}.

\bibitem{grossberg1980}
\bibinfo{author}{Grossberg, S.}
\newblock \bibinfo{title}{How does a brain build a cognitive code?}
\newblock \ERTdoi {10.1037/0033-295X.87.1.1} {,
  \emph{\bibinfo{journal}{Psychol. Rev.}} \textbf{\bibinfo{volume}{87}},
  \bibinfo{pages}{1--51} (\bibinfo{year}{1980})}.

\bibitem{fusi2005}
\bibinfo{author}{Fusi, S.}, \bibinfo{author}{Drew, P.~J.} \&
  \bibinfo{author}{Abbott, L.~F.}
\newblock \bibinfo{title}{Cascade models of synaptically stored memories}.
\newblock \ERTdoi {10.1016/j.neuron.2005.02.001} {,
  \emph{\bibinfo{journal}{Neuron}} \textbf{\bibinfo{volume}{45}},
  \bibinfo{pages}{599--611} (\bibinfo{year}{2005})}.

\bibitem{fusiabbott2007}
\bibinfo{author}{Fusi, S.} \& \bibinfo{author}{Abbott, L.~F.}
\newblock \bibinfo{title}{Limits on the memory storage capacity of bounded
  synapses}.
\newblock \ERTdoi {10.1038/nn1859} {, \emph{\bibinfo{journal}{Nat. Neurosci.}}
  \textbf{\bibinfo{volume}{10}}, \bibinfo{pages}{485--493}
  (\bibinfo{year}{2007})}.

\bibitem{bcm1982}
\bibinfo{author}{Bienenstock, E.~L.}, \bibinfo{author}{Cooper, L.~N.} \&
  \bibinfo{author}{Munro, P.~W.}
\newblock \bibinfo{title}{Theory for the development of neuron selectivity:
  orientation specificity and binocular interaction in visual cortex}.
\newblock \ERTdoi {10.1523/JNEUROSCI.02-01-00032.1982} {,
  \emph{\bibinfo{journal}{J. Neurosci.}} \textbf{\bibinfo{volume}{2}},
  \bibinfo{pages}{32--48} (\bibinfo{year}{1982})}.

\bibitem{vanrossum2000}
\bibinfo{author}{van Rossum, M. C.~W.}, \bibinfo{author}{Bi, G.~Q.} \&
  \bibinfo{author}{Turrigiano, G.~G.}
\newblock \bibinfo{title}{Stable {Hebbian} learning from spike timing-dependent
  plasticity}.
\newblock \ERTdoi {10.1523/JNEUROSCI.20-23-08812.2000} {,
  \emph{\bibinfo{journal}{J. Neurosci.}} \textbf{\bibinfo{volume}{20}},
  \bibinfo{pages}{8812--8821} (\bibinfo{year}{2000})}.

\bibitem{prezioso2016}
\bibinfo{author}{Prezioso, M.}, \bibinfo{author}{Merrikh~Bayat, F.},
  \bibinfo{author}{Hoskins, B.}, \bibinfo{author}{Likharev, K.} \&
  \bibinfo{author}{Strukov, D.}
\newblock \bibinfo{title}{Self-adaptive spike-time-dependent plasticity of
  metal-oxide memristors}.
\newblock \ERTdoi {10.1038/srep21331} {, \emph{\bibinfo{journal}{Sci. Rep.}}
  \textbf{\bibinfo{volume}{6}}, \bibinfo{pages}{21331} (\bibinfo{year}{2016})}.

\bibitem{rahman2023}
\bibinfo{author}{Rahman, M.}, \bibinfo{author}{Bose, S.} \&
  \bibinfo{author}{Chakrabartty, S.}
\newblock \bibinfo{title}{On-device synaptic memory consolidation using
  {Fowler--Nordheim} quantum-tunneling}.
\newblock \ERTdoi {10.3389/fnins.2022.1050585} {,
  \emph{\bibinfo{journal}{Front. Neurosci.}} \textbf{\bibinfo{volume}{16}},
  \bibinfo{pages}{1050585} (\bibinfo{year}{2023})}.

\bibitem{essex2026}
\bibinfo{author}{Essex, A.~E.}, \bibinfo{author}{Janson, N.~B.},
  \bibinfo{author}{Norris, R.~A.} \& \bibinfo{author}{Balanov, A.~G.}
\newblock \bibinfo{title}{Memorization and forgetting in a learning {Hopfield}
  neural network: bifurcation mechanisms, attractors, and basins}.
\newblock \ERTdoi {10.1063/5.0293682} {, \emph{\bibinfo{journal}{Chaos}}
  \textbf{\bibinfo{volume}{36}}, \bibinfo{pages}{083138}
  (\bibinfo{year}{2026})}.

\bibitem{eisenmann2023}
\bibinfo{author}{Eisenmann, L.}, \bibinfo{author}{Monfared, Z.},
  \bibinfo{author}{G{\"o}ring, N.} \& \bibinfo{author}{Durstewitz, D.}
\newblock \bibinfo{title}{Bifurcations and loss jumps in {RNN} training}.
\newblock \ERTdoi {10.52202/075280-3090} {, In \emph{\bibinfo{booktitle}{Adv.
  Neural Inf. Process. Syst.}}, vol.~\bibinfo{volume}{36},
  \bibinfo{pages}{70511--70547}} (\bibinfo{year}{2023}).

\bibitem{foradori2026}
\bibinfo{author}{Foradori, A.} \emph{et~al.}
\newblock \bibinfo{title}{Memory in integrated photonic neural networks: from
  physical mechanisms to neuromorphic architectures}.
\newblock \bibinfo{howpublished}{Preprint at
  \url{https://arxiv.org/abs/2604.22620}} (\bibinfo{year}{2026}).

\bibitem{schmidt2019}
\bibinfo{author}{Schmidt, D.} \emph{et~al.}
\newblock \bibinfo{title}{Tracking dark excitons with exciton polaritons in
  semiconductor microcavities}.
\newblock \ERTdoi {10.1103/PhysRevLett.122.047403} {,
  \emph{\bibinfo{journal}{Phys. Rev. Lett.}} \textbf{\bibinfo{volume}{122}},
  \bibinfo{pages}{047403} (\bibinfo{year}{2019})}.

\bibitem{kurdyubov2021}
\bibinfo{author}{Kurdyubov, A.~S.} \emph{et~al.}
\newblock \bibinfo{title}{Optical control of a dark exciton reservoir}.
\newblock \ERTdoi {10.1103/PhysRevB.104.035414} {,
  \emph{\bibinfo{journal}{Phys. Rev. B}} \textbf{\bibinfo{volume}{104}},
  \bibinfo{pages}{035414} (\bibinfo{year}{2021})}.

\bibitem{suchomel2017}
\bibinfo{author}{Suchomel, H.} \emph{et~al.}
\newblock \bibinfo{title}{Prototype of a bistable polariton field-effect
  transistor switch}.
\newblock \ERTdoi {10.1038/s41598-017-05277-1} {, \emph{\bibinfo{journal}{Sci.
  Rep.}} \textbf{\bibinfo{volume}{7}}, \bibinfo{pages}{5114}
  (\bibinfo{year}{2017})}.

\bibitem{opala2019}
\bibinfo{author}{Opala, A.}, \bibinfo{author}{Ghosh, S.},
  \bibinfo{author}{Liew, T. C.~H.} \& \bibinfo{author}{Matuszewski, M.}
\newblock \bibinfo{title}{Neuromorphic computing in {Ginzburg--Landau}
  polariton-lattice systems}.
\newblock \ERTdoi {10.1103/PhysRevApplied.11.064029} {,
  \emph{\bibinfo{journal}{Phys. Rev. Applied}} \textbf{\bibinfo{volume}{11}},
  \bibinfo{pages}{064029} (\bibinfo{year}{2019})}.

\bibitem{ballarini2020}
\bibinfo{author}{Ballarini, D.} \emph{et~al.}
\newblock \bibinfo{title}{Polaritonic neuromorphic computing outperforms linear
  classifiers}.
\newblock \ERTdoi {10.1021/acs.nanolett.0c00435} {,
  \emph{\bibinfo{journal}{Nano Lett.}} \textbf{\bibinfo{volume}{20}},
  \bibinfo{pages}{3506--3512} (\bibinfo{year}{2020})}.

\bibitem{takata2015reader}
\bibinfo{author}{Takata, K.}, \bibinfo{author}{Marandi, A.} \&
  \bibinfo{author}{Yamamoto, Y.}
\newblock \bibinfo{title}{Quantum correlation in degenerate optical parametric
  oscillators with mutual injections}.
\newblock \ERTdoi {10.1103/PhysRevA.92.043821} {, \emph{\bibinfo{journal}{Phys.
  Rev. A}} \textbf{\bibinfo{volume}{92}}, \bibinfo{pages}{043821}
  (\bibinfo{year}{2015})}.

\bibitem{wang2013opo}
\bibinfo{author}{Wang, Z.}, \bibinfo{author}{Marandi, A.},
  \bibinfo{author}{Wen, K.}, \bibinfo{author}{Byer, R.~L.} \&
  \bibinfo{author}{Yamamoto, Y.}
\newblock \bibinfo{title}{Coherent {Ising} machine based on degenerate optical
  parametric oscillators}.
\newblock \ERTdoi {10.1103/PhysRevA.88.063853} {, \emph{\bibinfo{journal}{Phys.
  Rev. A}} \textbf{\bibinfo{volume}{88}}, \bibinfo{pages}{063853}
  (\bibinfo{year}{2013})}.

\bibitem{wouters2007}
\bibinfo{author}{Wouters, M.} \& \bibinfo{author}{Carusotto, I.}
\newblock \bibinfo{title}{Excitations in a nonequilibrium {Bose--Einstein}
  condensate of exciton polaritons}.
\newblock \ERTdoi {10.1103/PhysRevLett.99.140402} {,
  \emph{\bibinfo{journal}{Phys. Rev. Lett.}} \textbf{\bibinfo{volume}{99}},
  \bibinfo{pages}{140402} (\bibinfo{year}{2007})}.

\bibitem{hamerly2015}
\bibinfo{author}{Hamerly, R.} \& \bibinfo{author}{Mabuchi, H.}
\newblock \bibinfo{title}{Quantum noise of free-carrier dispersion in
  semiconductor optical cavities}.
\newblock \ERTdoi {10.1103/PhysRevA.92.023819} {, \emph{\bibinfo{journal}{Phys.
  Rev. A}} \textbf{\bibinfo{volume}{92}}, \bibinfo{pages}{023819}
  (\bibinfo{year}{2015})}.

\bibitem{konthasinghe2017}
\bibinfo{author}{Konthasinghe, K.} \emph{et~al.}
\newblock \bibinfo{title}{Self-sustained photothermal oscillations in
  high-finesse {Fabry--Perot} microcavities}.
\newblock \ERTdoi {10.1103/PhysRevA.95.013826} {, \emph{\bibinfo{journal}{Phys.
  Rev. A}} \textbf{\bibinfo{volume}{95}}, \bibinfo{pages}{013826}
  (\bibinfo{year}{2017})}.

\bibitem{demirag2021}
\bibinfo{author}{Demira{\u g}, Y.} \emph{et~al.}
\newblock \bibinfo{title}{{PCM}-trace: scalable synaptic eligibility traces
  with resistivity drift of phase-change materials}.
\newblock \ERTdoi {10.1109/ISCAS51556.2021.9401446} {, In
  \emph{\bibinfo{booktitle}{2021 IEEE International Symposium on Circuits and
  Systems (ISCAS)}}, \bibinfo{pages}{1--5}} (\bibinfo{year}{2021}).

\bibitem{ERTdeposit}
\bibinfo{author}{Berloff, N.~G.}
\newblock \bibinfo{title}{Code, data and figure sources for ``{M}atter that
  learns what matters''}.
\newblock \bibinfo{howpublished}{Zenodo
  \url{https://doi.org/10.5281/zenodo.23775077}} (\bibinfo{year}{2026}).

\end{thebibliography}
\end{document}